\documentclass{cernyrep}
\usepackage[T1]{fontenc}
\usepackage[bookmarks, colorlinks=true, linktoc=page, linkcolor=black, citecolor=black, urlcolor=blue]{hyperref}
\usepackage{fancyhdr}
\fancyhfoffset{4 mm}
\fancypagestyle{ARTTITLE}{%
\fancyhf{} 
\lhead{\small{Proceedings of the 2018 CERN--Accelerator--School course on \it{Beam Instrumentation}, Tuusula, (Finland)}} 
\lfoot{Available online at \url{https://cas.web.cern.ch/previous-schools}}
\rfoot{\thepage\hspace*{3mm}}
 
}

\usepackage{varwidth}
\usepackage{xcolor}

\begin{document}

\title{Diagnostics Examples from CTF3}
\author{F. Tecker}
\institute{CERN, Geneva, Switzerland}

\begin{abstract}
After a short introduction of CLIC, the Compact Linear Collider, and its test facility CTF3 (CLIC Test Facility 3), this paper gives an overview and some examples of the diagnostics used at CTF3. 
\end{abstract}
\maketitle
\thispagestyle{ARTTITLE}

\section{Introduction}
After the discovery of the Higgs particle, there is large interest to study its properties in much more detail. A high-energy lepton collider would offer the potential for precision physics due to the 'cleaner' environment in the collisions with much less perturbing background events.

The highest energy lepton collider so far was the LEP collider, which was installed in the present LHC tunnel. It accelerated electrons and positrons to a centre-of-mass energy up to slightly above 200~GeV. LEP was being limited in energy by synchrotron radiation, as the RF system had to provide sufficient power to replace the energy loss that amounted 3\% per turn at the highest energy. The emitted power scales with the fourth power of the energy and with the inverse of the square of the radius, so a future ring collider will require a much larger radius. A study is ongoing for such a machine, the Future Circular Collider (FCC)~\cite{ref:FCC}, that would require a 80-100\ km circumference tunnel.

Another option is to avoid the bending that leads to the emission of synchrotron radiation and build a linear collider with two opposing linear accelerators for electrons and positrons. Two different projects pursue this approach, the International Linear Collider (ILC)~\cite{ref:ILC} and the Compact Linear Collider (CLIC)~\cite{CLIC, CLIC-CDR}.

\subsection{Compact Linear Collider --- CLIC}
\begin{figure}[!tb]
\begin{center}
\includegraphics[width=\textwidth]{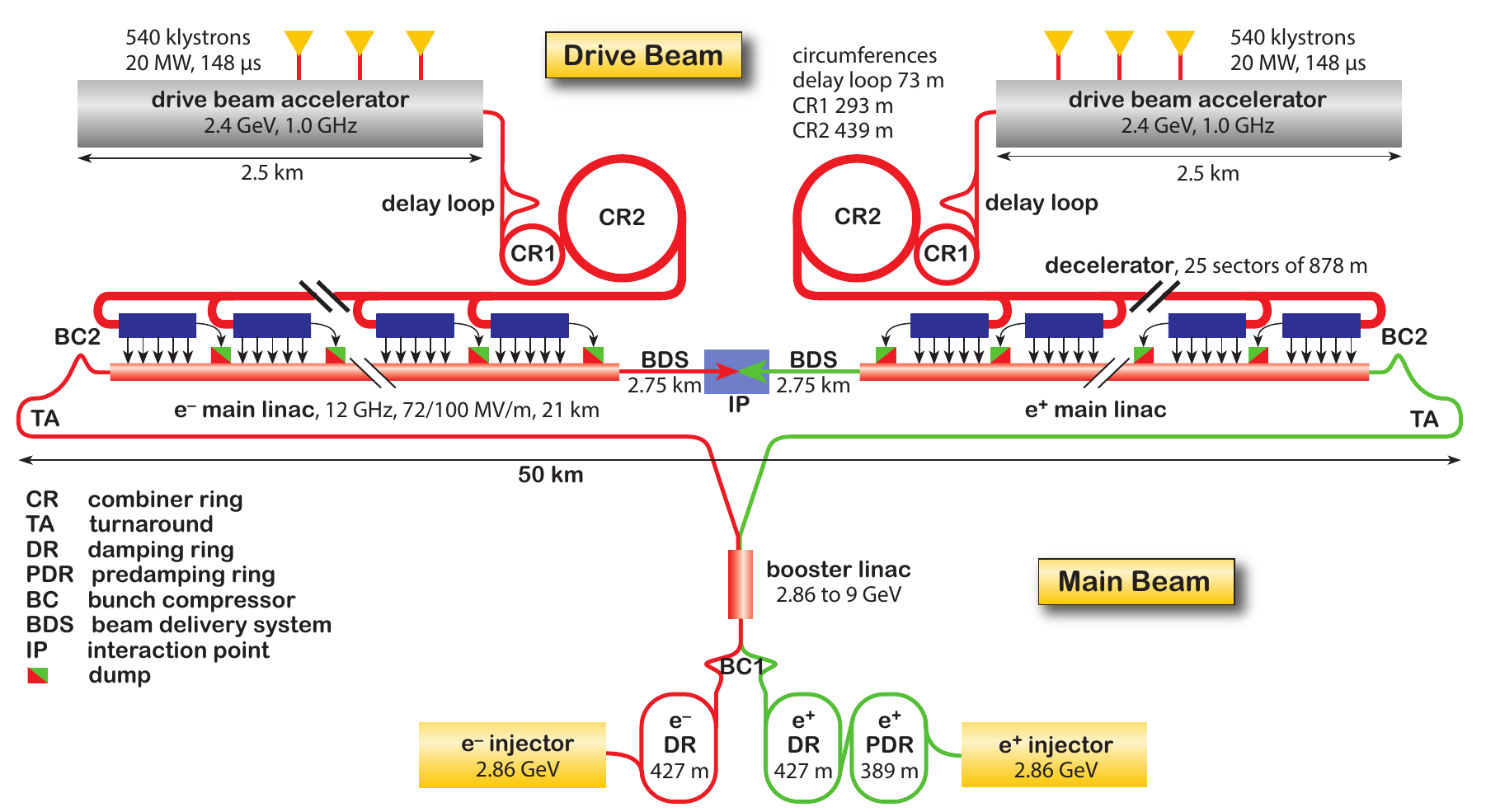}
\caption{\label{fig:CLIClayout}CLIC overall layout for a centre-of-mass energy of 3~TeV}
\end{center}
\end{figure}

CLIC is a TeV-scale high-luminosity linear e+e- collider under development at CERN. CLIC is foreseen to be built in stages, at centre-of-mass energies of 380~GeV, 1.5~TeV and 3~TeV, respectively, with a site length ranging from 11 km to 50 km, for an optimal exploitation of its physics potential.
CLIC relies on a two-beam acceleration scheme, where normal-conducting high-gradient
12~GHz accelerating structures are powered via a high-current drive beam. The accelerating gradient reaches 100~MV/m with an RF frequency of 12~GHz. The layout is shown in Fig.~\ref{fig:CLIClayout}.

The Drive Beam Generation is an essential part of the CLIC design. It requires the efficient generation of a high-current electron beam with the time structure needed to generate 12~GHz RF power. CLIC relies on a novel scheme of fully loaded acceleration in normal conducting travelling wave structures, followed by beam current and bunch frequency multiplication by funnelling techniques in a series of delay lines and rings, using injection by RF deflectors. All the combination process must be very well controlled in terms of transverse and longitudinal beam parameters in order to obtain the required transverse emittance, bunch length, and bunch distance of the drive beam.

\subsection{CTF3 --- CLIC Test Facility 3}
\begin{figure}[!b]
\begin{center}
\includegraphics[width=\textwidth]{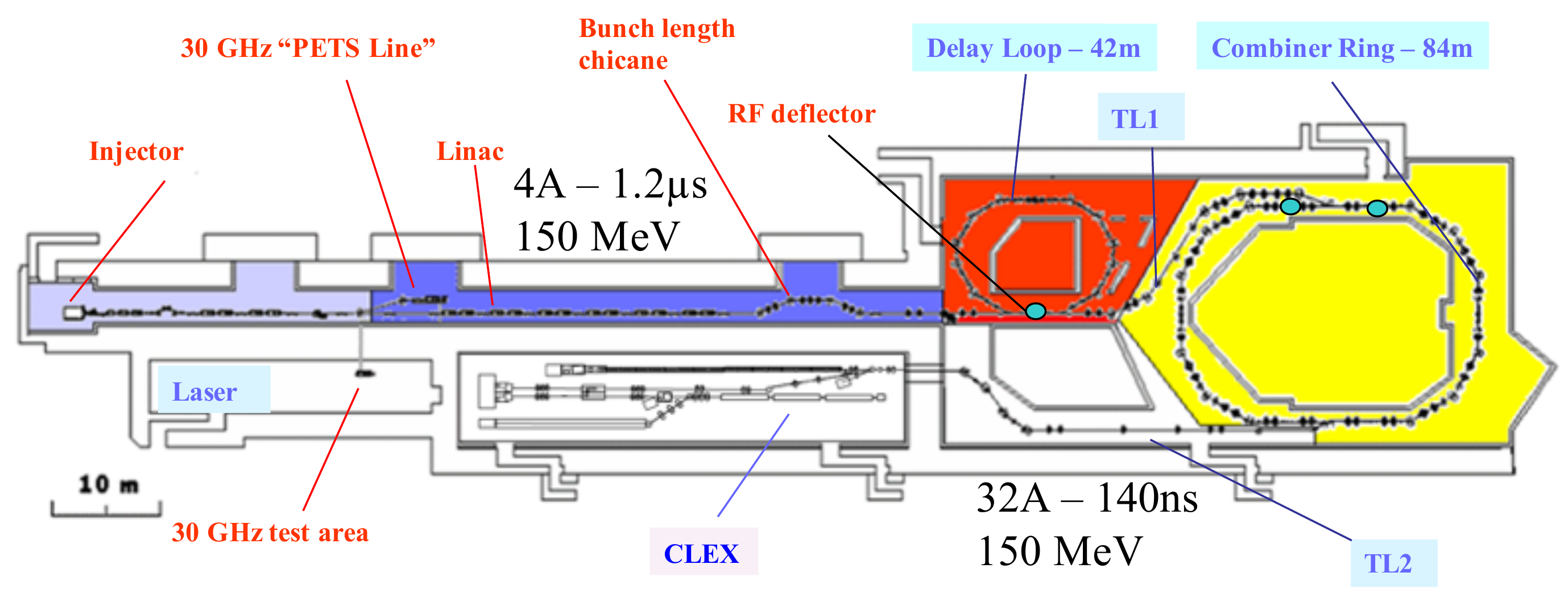}
\caption{\label{fig:layout}CTF3 overall layout}
\end{center}
\end{figure}

The aim of CTF3~\cite{CTF3-DR} (see Fig.~\ref{fig:layout}) was to prove the main feasibility issues of the CLIC two-beam acceleration technology. The two main points which CTF3 demonstrated were the generation of a very high current drive beam and its use to efficiently produce and transfer RF power to high-gradient accelerating structures, used to bring the main beam to high energies.

CTF3 consisted of a 150~MeV electron linac followed by a 42~m long Delay Loop (DL) and a 84~m Combiner Ring (CR). The beam current was first doubled in the delay loop and then multiplied again by a factor four in the combiner ring by interleaving bunches using transverse deflecting RF cavities.
The drive beam could be sent to an experimental area (CLEX) to be used for deceleration and to produce RF power for two-beam experiments.
In CLEX, one line (Test Beam Line, TBL) was used to decelerate the drive beam in a string of power generating structures (PETS). The drive beam could alternatively be sent in another beam line (Two-Beam Test Stand, TBTS), where a PETS is used to power one or more structures, used to further accelerate a 200~MeV electron beam provided by a dedicated injector (CALIFES). Table~\ref{tab:parameters} shows some of the beam parameters of CTF3 compared to CLIC.

\begin{table}[!tb]
\begin{center}
\begin{tabular}[b]{p{1.31in}p{0.7in}p{0.65in}p{1.65in}}
 \hline 
\multirow{2}*{\bf Parameter} & \multirow{2}*{\bf Unit} & \bf CLIC\newline nominal & \multirow{2}*{\bf CTF3} \\
\hline 

Initial beam current & A & 4.2 & 4 \\ 
Final beam current& A & 100 & 28  \\
Bunch charge $Q_{\rm b}$ & nC & 8.4 & 4 (2.3 nom.)  \\ 
Emittance, \newline norm rms & $\pi$\,mm\,mrad & $\leq$ 150 & $\leq$ 250/100 (factor 8 comb. beam, horizontal/vertical)  \\ 
Bunch length & mm & $\leq$ 1 & $\leq$ 1 \\
Energy $E$ & GeV & 2.4 & 0.120 \\
T${}_\text{pulse}$ initial & $\mu$s & 140 & 1.4 \\
T${}_\text{pulse}$ final & ns & 240 & 140 (280) \\
Beam Load. Eff. & \% & 97 & 95 \\ \hline 
\end{tabular}
\caption{\label{tab:parameters}
CLIC nominal parameters compared to the CLIC Test Facility CTF3
}
\end{center}
\end{table}

\section{CTF3 Diagnostics}

Diagnostics in the CTF3 complex have primarily been designed for CTF3 beam parameters and used to commission and optimize the performance of the CTF3 machine~\cite{bib:CTF3_inst}. However, since the CTF3 drive beam is a small scale version of the proposed CLIC drive beam, it was natural to use the CTF3 machine 
as a test environment for CLIC-type beam diagnostics.  A large fraction of the non-interceptive CTF3 instruments, such as the beam position monitors, beam loss monitors and longitudinal beam monitors, can be adapted for  the CLIC drive beam parameters. However, because of the considerably higher bunch charge, higher beam energy and repetition rate in the CLIC drive beam, the CTF3 interceptive beam diagnostics, which are typically used for  providing transverse beam size measurements (emittance and energy spread), would be of limited use for the CLIC drive beam.

The following devices were installed in CTF3:
\begin{itemize}
\item Large variety of Beam Position Monitors (BPM)
\begin{itemize}
\item High resolution cavity BPM
\item Inductive pick-up
\item Strip-line BPM
\end{itemize}

\item Screens (OTR, fluorescence) for beam imaging

\item Several technologies of Beam Loss Monitors (BLM)

\item Longitudinal profile measurement with an RF deflecting cavity, streak camera, and electro-optical monitors

\item Fast Wall Current Monitors

\item Segmented dump for time resolved beam energy profile

\item mm-wave detectors for bunch length/spacing measurements.

\end{itemize}

\subsection{Diagnostics for position measurements}
There were 137 beam position monitors used in the CTF3 complex. The most abundant type, was the inductive pick-up that had been developed especially for CTF3~\cite{bib:BPM, bib:BPM2}. The pickup detects the beam image current circulating on the vacuum chamber using eight electrodes. Several versions of this type of pick-up were designed, built and installed to match the size and shape of the vacuum chamber in the different parts of the CTF3 complex~\cite{bib:BPM2, bib:BPM_Frascati, bib:bps}.
More details on the beam position monitors for CLIC can be found in \cite{bib:CTF3_inst}.  

\subsection{Transverse profile measurements}
CTF3 had 21 operational optical transition radiation (OTR) based TV stations, in order to measure the transverse profile of the beam throughout the complex. 
The imaging system could provide a resolution of up to 20\,$\rm{\mu m}$, and could provide time resolved transverse beam profiles
when coupled to intensified gated cameras. 
Some of the OTR based TV stations installed in CTF3 were equipped with two screens for low and high charge operation respectively, a calibration target 
and a replacement chamber to ensure a continuity of the beam line, when the device was not in use, thus minimising wakefield effects~\cite{bib:CTF3_inst}. 

\begin{figure}[!h]
\centering
\includegraphics[width=0.8\textwidth]{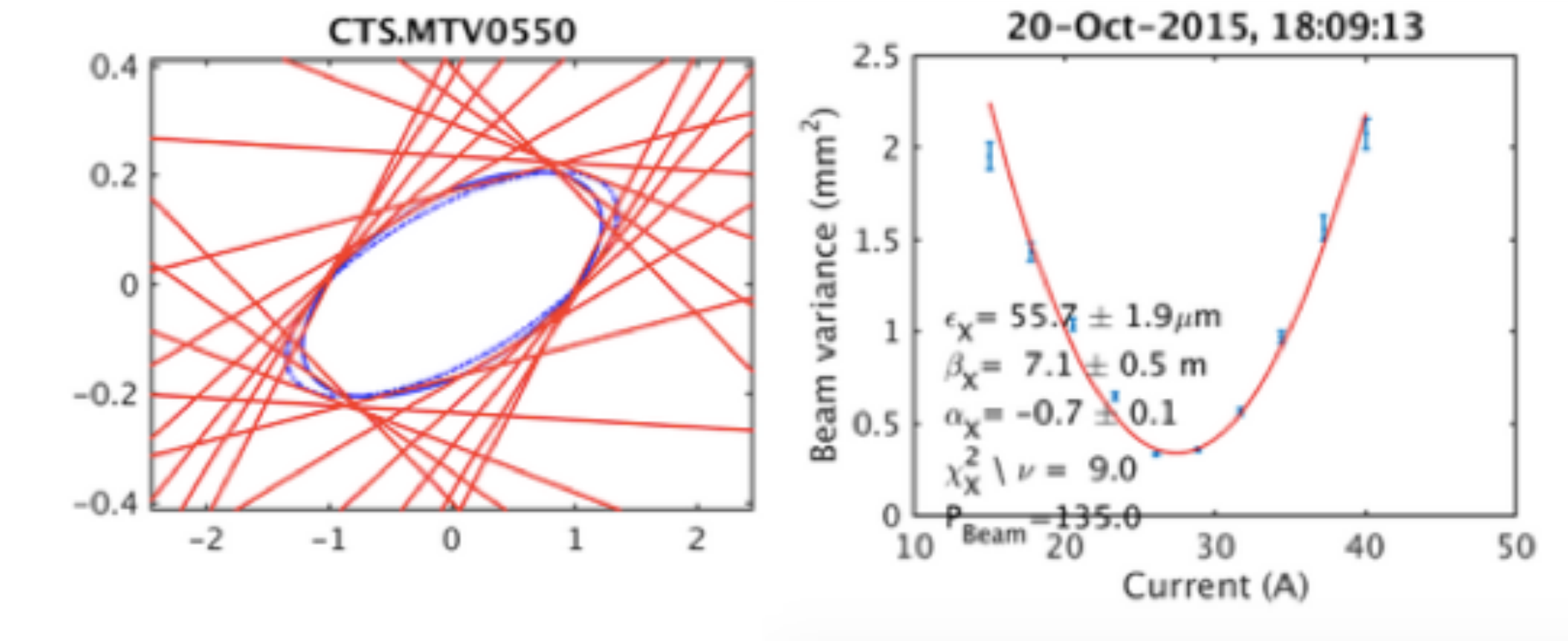}
 \caption{\label{fig:quadscan}Quadrupole scan example. The left plot shows the phase space with the projected beam size measurements (red lines) and the fitted matched beam ellipse (blue). The right plot shows the beam variance as a function of the quadrupole current and the model fit with the calculated beam parameters.}
\end{figure}

The screens were used in spectrometer lines to measure the average energy spread, and with 
the ``quadrupole scan" technique in regions of minimal dispersion to measure the beam emittance. The current in upstream quadrupoles is varied, which changes the betatron phase advance to the screen, so the betatron phase space projection changes. Using the known transfer matrices for the given quadrupole strength, the initial transverse beam parameters (Twiss parameters $\beta$, $\alpha$, transverse emittance $\varepsilon$) can be extracted from a fit to the data. Fig.~\ref{fig:quadscan} shows an example of a quadrupole scan.   

\begin{figure}[!h]
\centering
\includegraphics[width=0.9\textwidth]{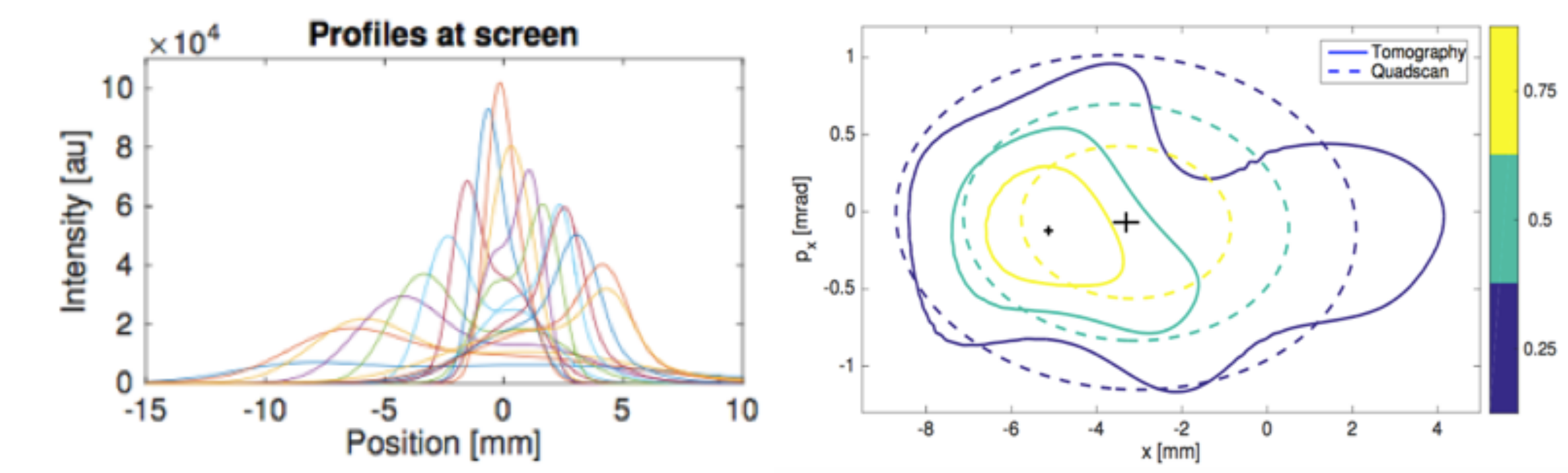}
 \caption{\label{fig:quadscantomo}Example for a quadrupole scan tomography. The left plot shows projected profiles for different quadrupole settings. The right plot shows the tomography reconstruction of the phase space density contours together with the conventional quadrupole scan contours.}
\end{figure}

Using the full profile data, it was also possible to perform a tomography and reconstruct the phase space distribution by applying an inverse Radon transform. An example is given in Fig.~\ref{fig:quadscantomo}.

\subsection{Diagnostics for time resolved energy measurements}
\begin{figure}[!hbt]
\centering
 \includegraphics[width=\textwidth]{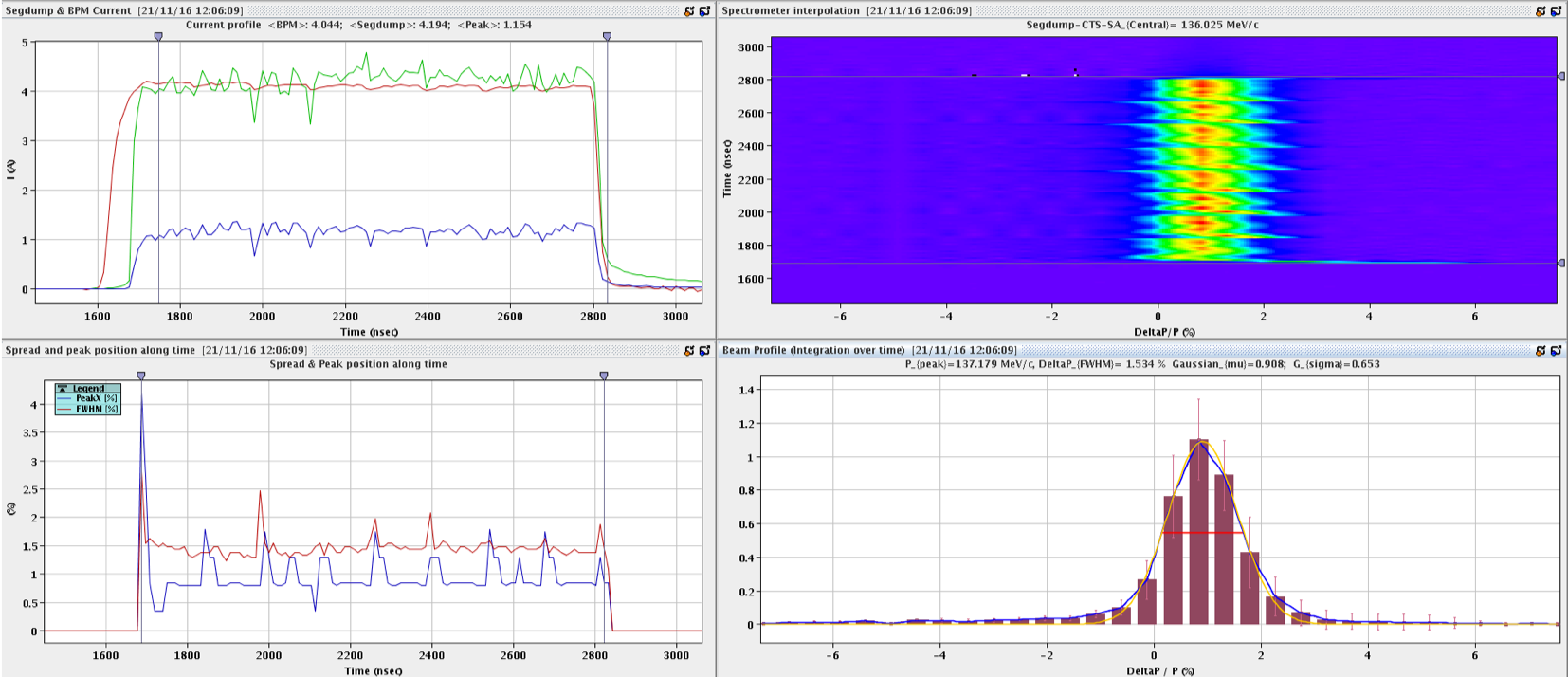}
\caption{\label{fig:segdump}Example of acquisition software analysis for the segmented dump. The plots show the intensity along the beam pulse (top left), mean momentum and momentum spread along the pulse (bottom left), the momentum distribution along the pulse (top right), and the integrated momentum distribution (bottom right). The repetitive time pattern in the momentum distribution results from the phase switching with 140~ns period in the injector required for the bunch recombination.}
\end{figure}

In order to optimize the overall efficiency of the drive beam generation, the accelerating structures~\cite{bib:sica} 
are operated in a fully beam-loaded condition, meaning that all the RF power, except for ohmic losses, is transferred into beam energy. In this mode of operation, the resulting energy spectrum shows a strong transient behaviour, with higher energies in the first $10-50~\text{ns}$ of the pulse. Also any current variation during the pulse translates into an energy variation. Time-resolved spectrometry was therefore an essential beam diagnostic to correctly tune the timing of the rf pulses powering the accelerating structures in the linac. For this purpose segmented beam dumps~\cite{bib:SegDump, bib:SegTBL} or multi-anode photomultiplier (MAPMT)~\cite{bib:SegPMT} coupled to an OTR screen, were developed and were used for the daily optimization of the CTF3 Linac. The segmented dump was built from 32~Tungsten plates, 2~mm thick each, spaced by 1~mm. A multi-slit collimator was installed just upstream of the segmented dump, in order to limit the deposited energy in the dump plates. The deposited charge was directly read with 50~$\Omega$ impedance to the ground, with a time resolution better than 10~ns, limited by the bandwidth of the installed digitizers.
The channels were fully cross-calibrated by sweeping the beam over the detector. This tool allowed injector and linac optimization to reach a momentum spread of $\sigma_p/p = 0.7\%$. An example of the optimized beam pulse is shown in Fig.~\ref{fig:segdump}.  

\subsection{Longitudinal beam diagnostics}
Longitudinal beam manipulation in the CTF3 and the CLIC drive beam are similar. Bunch length manipulation and bunch frequency multiplication has led to the development of adequate non-intercepting devices based on the detection of optical photons by a streak camera~\cite{bib:welsch, bib:welsch2} and on radio frequency pick-ups~\cite{bib:Long_2010}.

Another technique for longitudinal beam profile measurements, which already demonstrated extremely good time resolution ($10~\text{fs}$)~\cite{bib:RF_def_Emma}, is based on RF deflecting cavities. Such cavities were used in CTF3 for the RF injection in the Delay Loop and Combiner Ring and could also serve for this kind of bunch length measurements~\cite{bib:RF_def_Ghigo}. 

\subsubsection{Streak Camera}
The streak camera was used for bunch length measurements but primarily as an essential instrument to achieve the correct time structure of the combined beam downstream the Delay Loop and the Combiner Ring. The light from several of the OTR screens could be sent to a streak camera. The synchrotron radiation at the end-of-linac chicane, the Delay Loop, and Combiner Ring could also be transported to a streak camera. 

\begin{figure}[!b]
  \centering
 a)
 \includegraphics[width=0.47\textwidth]{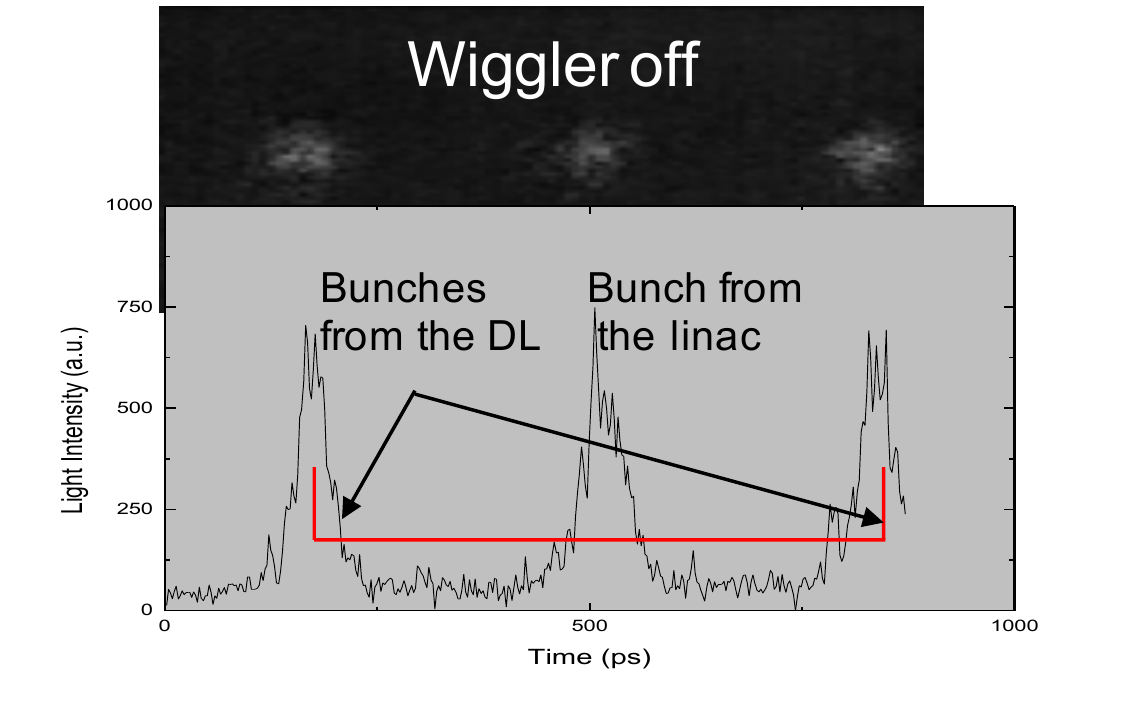}
 \hfill
 b)
 \includegraphics[width=0.47\textwidth]{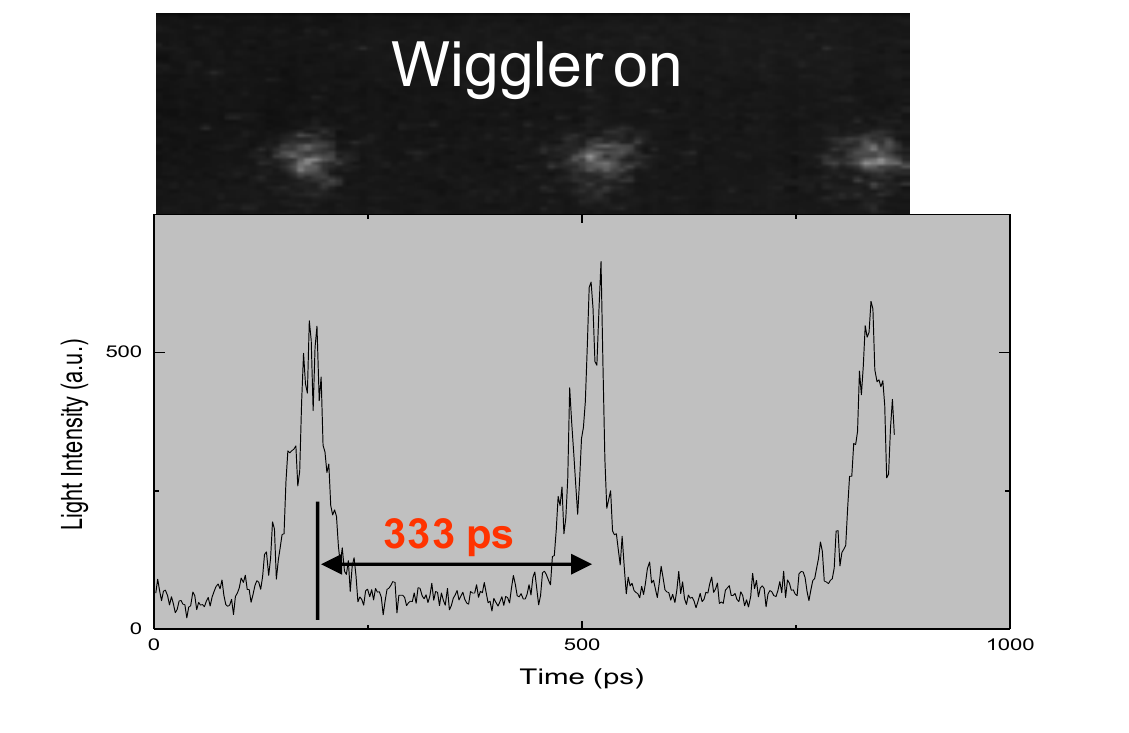}

 \caption{\label{fig:sc-dl}Streak camera measurement downstream the Delay Loop for two configurations of the magnetic wiggler for path-length adjustment. The plots show the streak camera image and the projection on the time axis. a) Without wiggler, the bunches from the Delay Loop arrive 12~ps (3.6~mm) before the ideal time (position) compared to the linac bunches. b) The wiggler was adjusted to the correct current to have Delay Loop and linac bunches equidistant.}
\end{figure}

Wiggler magnets in both Delay Loop and Combiner Ring could tune the path-length in these rings. This allowed the adjustment of the time of flight such that the bunches through the Delay Loop and straight from the linac were precisely interleaved in between each other (see Fig.~\ref{fig:sc-dl}).

\begin{figure}[!b]
  \centering
 \includegraphics[width=\textwidth]{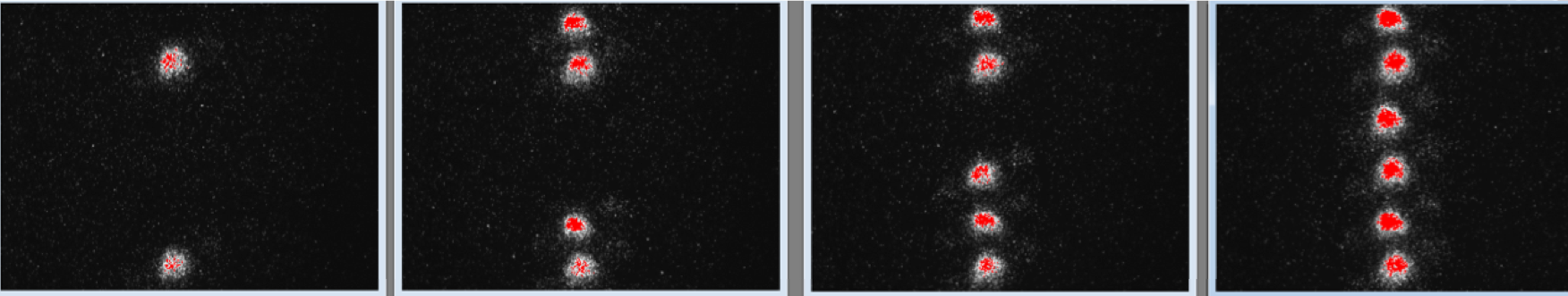}
 \caption{\label{fig:sc-recombination} Example of streak camera images of the bunch train recombination in the Combiner Ring for a recombination factor four. The time axis points in the vertical direction. The left image shows two consecutive bunches of the first train. The following images show the bunches of the following trains interleaved. }
\end{figure}

Similarly in the Combiner Ring, the path-length was tuned by the wiggler 
to inject the bunches with equidistant spacing. The adjustment range of the wiggler was large enough to operate with combination factors four and five. Fig.~\ref{fig:sc-recombination} shows an example of streak camera acquisitions at different times during the recombination.

The streak camera has also been used in order to study the phase
coding of the bunches, which is introduced by a sub-harmonic bunching (SHB) system at the beginning of the linac.
The measurements have shown that the phase switch is
performed in 6~ns with a presence of 8\% satellite bunches
at 3~GHz~\cite{bib:SHB}.

\subsubsection{RF Pickups}
Several single-waveguide pickups were installed along the machine~\cite{bib:freqmult}.
The WR28 waveguide used acts as a high-pass filter for the beam-induced signal. Shorter bunches contain a spectrum extending to higher frequencies, resulting in a larger signal for the same beam charge.
The power is detected by a fast Schottky barrier detector, sensitive in the 26.5-40~GHz range, and digitised.
The qualitative signal was used for everyday optimization of the accelerator.  

\begin{figure}[!b]
\begin{minipage}{0.45\textwidth}
 \includegraphics[width=\textwidth]{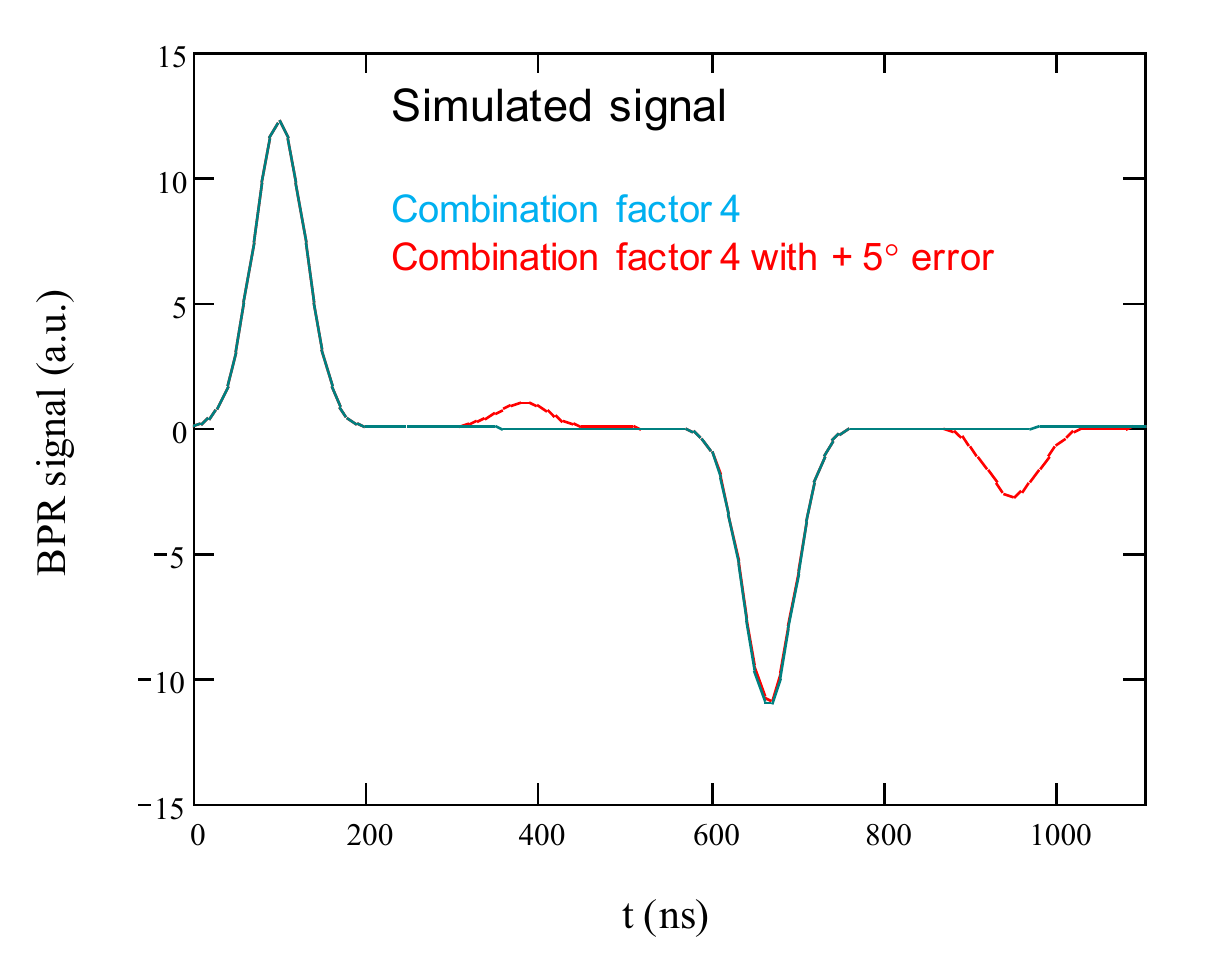}
\end{minipage}
\begin{minipage}{0.45\textwidth}
  \includegraphics[width=\textwidth]{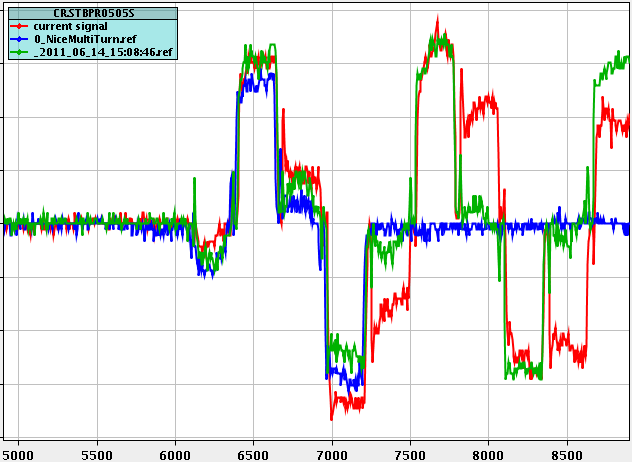}
\end{minipage}
 \caption{\label{fig:CR-BPR}Left: Simulated signal of the CR pickup mixed with the 3~GHz RF reference for 4 turns with a factor 4 recombination. Since the phase advance per 240~ns revolution period is $90^\degree$ for the correct path-length, the signal alternates between maximum (positive and negative) amplitude and zero. A path-length error becomes clearly visible as the phase difference adds up for each turn. (Courtesy: R.~Corsini);
 Right: Measured signal for a $\approx$270~ns long beam pulse. The horizontal axis is in ns, the vertical in arbitrary units. The green trace shows 10 turns of a beam pulse with well adjusted path-length, the red trace reveals a path-length error as a phase slippage.}
\end{figure}

The signal of a button pickup in the Combiner Ring was mixed with a 3~GHz RF reference signal with adjustable phase. The Combiner Ring path-length for a factor 4 combination is $(N \pm 1/4) \lambda$ (with $N$ integer, $\lambda$: RF wavelength of the 3~GHz RF deflector), so the phase advance per turn of the bunches with respect to the reference RF is $90^\degree$. For the correct path-length, the reference phase can be adjusted such that the signal will alternate between a maximum and a zero signal for the successive turns, as shown in Fig.~\ref{fig:CR-BPR}.
Even a small path-length error of a few degrees becomes clearly visible after a number of turns.
This signal was used in the daily operation to monitor and optimize the path-length in the CR, as it was always available in the operation, compared to streak camera measurements, that every time needed setup by a specialist. 

\begin{figure}[!tb]
 \centering
 \includegraphics[width=0.55\textwidth]{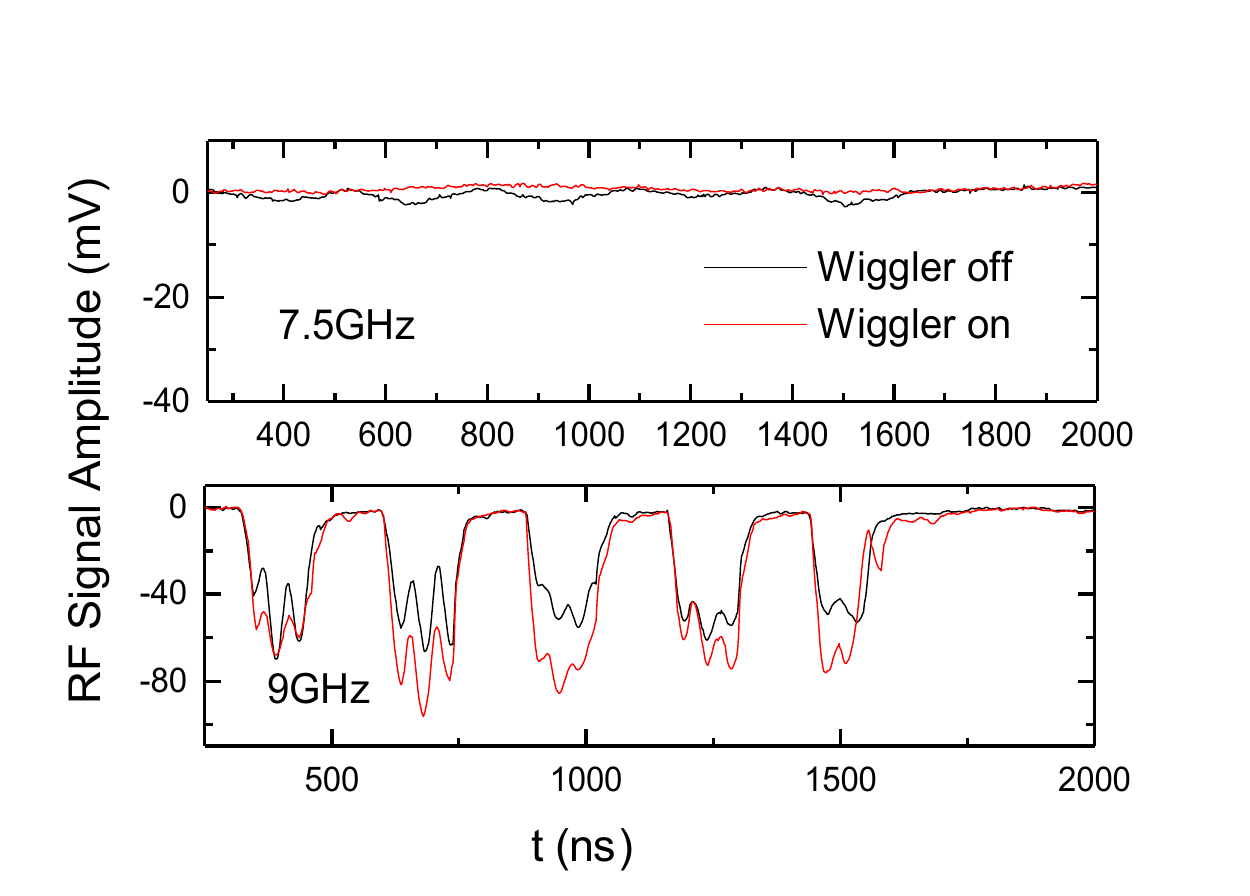}
 \caption{\label{fig:DL-path}Signals of the RF pickup downstream the DL for the 7.5 and 9~GHz channels. The wiggler was optimized for the correct path-length. In this case (red trace), the 7.5~GHz component is close to zero and the 9~GHz signal higher, compared to the case with the wiggler off (black trace) where the path-length is not optimal.}
\end{figure}

Another RF pickup used a different method for the measurement of the bunch-train recombination quality. A monitor signal downstream the Delay Loop was split into several channels with bandpass filters centred at frequencies of 7.5, 9, 10.5, and 12~GHz. Since the initial beam has a 1.5~GHz bunch repetition frequency, all harmonic frequency channels will measure a signal. After the DL recombination, ideally only harmonics of 3~Ghz should show a signal. An error in the path-length shows in the odd 1.5~Ghz harmonics. Fig.~\ref{fig:DL-path} shows an example.

A similar setup was installed in the CR with filters of 6, 9, 12, and 15~GHz. For the ideal combination, only the 12~Ghz component should show a signal at the end of the combination.

Finally, another installation, a microwave spectrometer, was used for bunch length monitoring~\cite{bib:mwspectro}. The RF signal from the beam was sent on a series of reflecting low pass filters, in order to distribute it into four bands in frequency ranges from 30 to 170~Ghz. Each detection station had two down-mixing stages. The setup has the advantage that it is non-intercepting but it is sensitive to beam position and current variations. 
 
\subsubsection{RF Deflector}
\begin{figure}[!bt]
\begin{minipage}{\textwidth}
  \centering
	\includegraphics[width=0.6\textwidth]{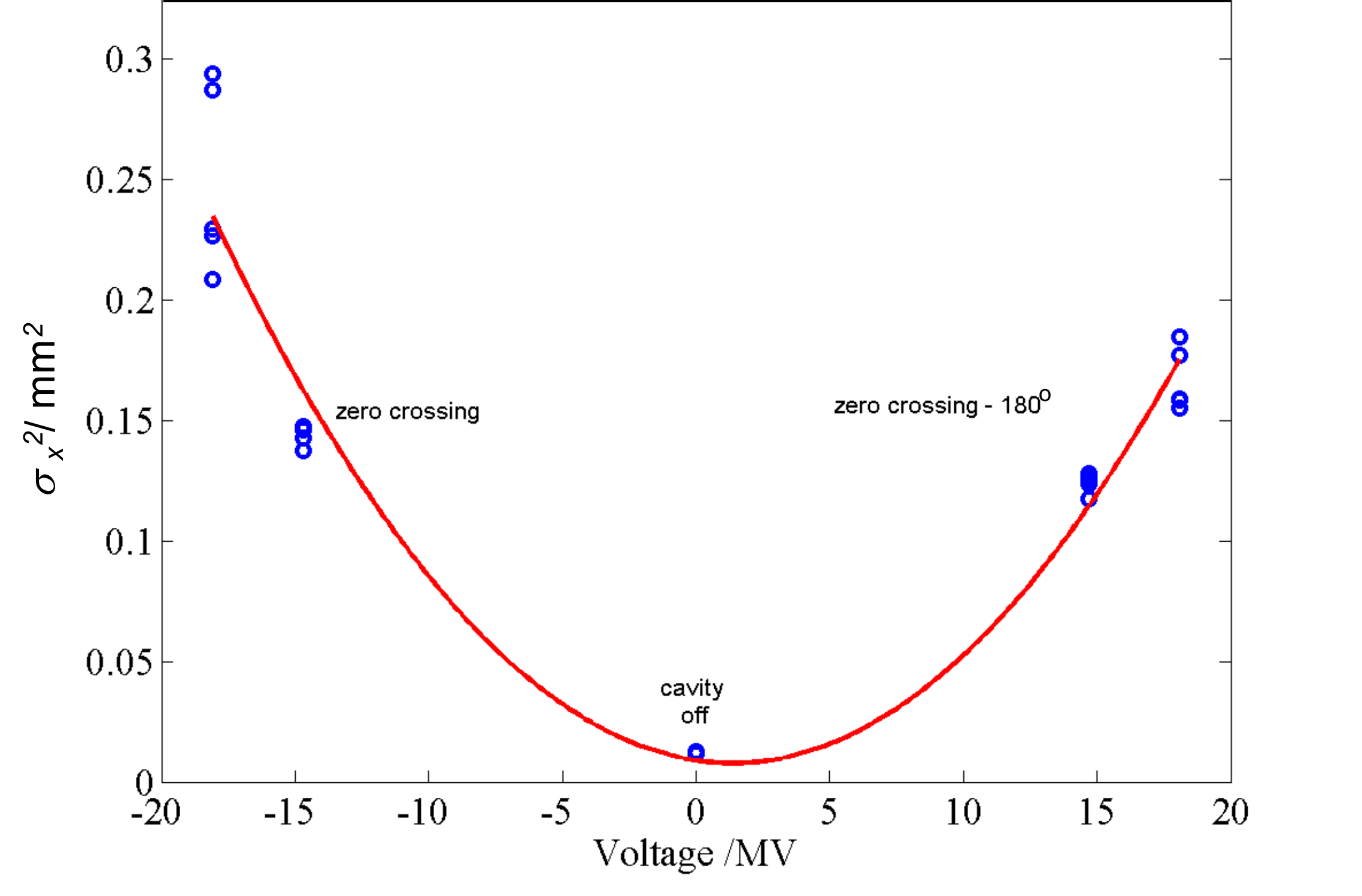}
\end{minipage}
 \caption{\label{fig:defl-scan}Beam size variation as a function of the field in the RF deflector.}
\end{figure}

The RF deflector for the bunch train combination was also used for a dedicated measurement of the bunch length. This uses the time-varying deflecting field of the cavity near the RF zero-crossing to transform the the time information into a spacial information. The bunch length can be deduced measuring the beam size at a downstream screen with the RF deflector powered and not powered. The measured transverse beam size $\sigma_x$ at the screen is given by
\begin{equation}
\sigma_x = \sqrt{\sigma_{x0}^2 + \sigma_z^2 \beta_c \beta_p 
\left( \frac{2\pi e V_0}{\lambda_\mathrm{rf} E_0} \sin\Delta\Psi_x
\cos\varphi_\mathrm{rf} \right)^2 }
\end{equation} 
where $\sigma_{x0}$ is the beam size on the screen without deflection, $\sigma_z$ the bunch length, $\beta_c$ and $\beta_s$ the beta function at the deflector and screen, respectively, $V_0$ the deflecting voltage, 
$\lambda_\mathrm{rf}$ the RF deflector wavelength, $E_0$ the beam energy,
$\Delta\Psi_x$ the betatron phase advance between deflector and screen, and $\varphi_\mathrm{rf}$ the RF deflector phase. Since the deflecting voltage $V_0$ was not known precisely, the RF deflection angle was calibrated using a BPM close to the screen and the beam transfer matrix between the cavity and the BPM.
Fig.~\ref{fig:defl-scan} shows an example of a measurement with different RF voltages for both negative and positive zero crossing.
The bunch length can be determined from the results of the beam size fit.

\section{Conclusion}
The CLIC Test Facility CTF3 had a lot of interesting diagnostics equipment. A good diagnostic is absolutely essential to optimize the performance, and only the variety of the instruments made it possible to reach the goals of CTF3 and show the feasibility of the CLIC Drive Beam generation.

Many other instruments in view of CLIC were tested at CTF3 but it was not possible to cover all this within the lecture.

\section*{Acknowledgement}
I would like to express a big thanks to all my colleagues from CTF3,
in particular to Thibaut Lefevre, Roberto Corsini, Piotr Skowronski, Davide Gamba, Lars S{\o}by, Alexandra Andersson, and Anne Dabrowski.

\end{document}